\documentclass[conference]{IEEEtran}

\usepackage{graphicx}
\usepackage{color}
\usepackage{soul}
\definecolor{light-gray}{gray}{0.85}

\usepackage{cite} % Make references as [1-4], not [1,2,3,4]
\usepackage{url}  % Formatting web addresses
\usepackage{ifthen}  % Conditional
\usepackage{multicol}   %Columns package
\usepackage{graphicx,courier,amssymb,amsmath,times,caption}
\usepackage{xcolor}
\usepackage{epsfig,psfrag}
\usepackage{bm,graphicx,hhline,amsmath,amssymb,array}
\usepackage{rotating}
\usepackage{amsfonts}
\usepackage{textcomp}
\usepackage{upgreek}
\usepackage{comment}
\usepackage{subfig} %subfigure package
\usepackage{bbold}

\usepackage{floatrow}
\usepackage[acronyms]{glossaries}
\glsdisablehyper
\usepackage{tablefootnote}
\usepackage[linesnumbered,algoruled,boxed,lined]{algorithm2e}

\usepackage{pgfplots}
\pgfplotsset{compat=1.17}

\begin{document}

\title{Next Generation LoRaWAN: Integrating Multi-Hop Communications at 2.4 GHz}

\author{\IEEEauthorblockN{Riccardo Marini, Giampaolo Cuozzo}\\
\textit{CNIT, National Laboratory of Wireless Communications (WiLab)}, Bologna, Italy \\
\{riccardo.marini, giampaolo.cuozzo\}@wilab.cnit.it}

\maketitle

\begin{abstract}
The Internet of Things (IoT) revolution demands scalable, energy-efficient communication protocols supporting widespread device deployments. The LoRa technology, coupled with the LoRaWAN protocol, has emerged as a leading Low Power Wide Area Network (LPWAN) solution, traditionally leveraging sub-GHz frequency bands for reliable long-range communication. However, these bands face constraints such as limited data rates and strict duty cycle regulations. Recent advancements have introduced the 2.4 GHz spectrum, offering superior data rates and unrestricted transmission opportunities at the cost of reduced coverage and severe interference. To solve this trade-off, this paper proposes a novel hybrid approach integrating multi-band (i.e., sub-GHz and 2.4 GHz) and multi-hop communication into LoRaWAN, while preserving compatibility with the existing standard. The proposed network architecture retains Gateways (GWs) and End Devices (EDs) operating within the sub-GHz frequency while introducing multi-band Relays (RLs) that act as forwarding nodes for 2.4 GHz EDs.  Utilizing our previously developed open-source and standards-compliant simulation framework, we evaluate the network performance of our solution under realistic deployment scenarios. The results demonstrate substantial improvements compared to standard single-band and single-hop LoRaWAN networks, demonstrating the potential of this approach to redefine LPWAN capabilities and bridge the gap between current solutions and next-generation IoT applications.

\end{abstract}


%%%%%%%%%%%%%%%%%%%%%%%%%%%%%%%%%%%%% Chapter files
%%%%%%%%%%%%%%%%%%%%%%%%%%%%%%%%%%%%%%%%%%
%ho messo in ordine alfabetico
\newacronym{ack}{ACK}{Acknowledgment}
\newacronym{adr}{ADR}{Adaptive Data Rate}
\newacronym{bw}{BW}{Bandwidth}
\newacronym{cr}{CR}{Coding Rate}
\newacronym{crc}{CRC}{Cyclic Redundancy Check}
\newacronym{csma}{CSMA}{Carrier Sense Multiple Access}
\newacronym{ed}{ED}{End Device}
\newacronym{fec}{FEC}{Forward Error Correction}
\newacronym{gw}{GW}{Gateway}
\newacronym{ism}{ISM}{Industrial, Scientific and Medical}
\newacronym{iot}{IoT}{Internet of Thing}
\newacronym{lpwan}{LPWAN}{Low Power Wide Area Network}
\newacronym{mac}{MAC}{Medium Access Control}
\newacronym{ns}{NS}{Network Server}
\newacronym{pdu}{PDU}{Protocol Data Unit}
\newacronym{rx1}{RX1}{Receive Window 1}
\newacronym{rx2}{RX2}{Receive Window 2}
\newacronym{rt}{RT}{Router}
\newacronym{rl}{RL}{Relay}
\newacronym{sf}{SF}{Spreading Factor}
\newacronym{toa}{ToA}{Time On Air}
\newacronym{3gpp}{3GPP}{3rd Generation Partnership Project}
\newacronym{los}{LOS}{Line-of-Sight}
\newacronym{nlos}{NLOS}{Non-Line-of-Sight}
\newacronym{rssi}{RSSI}{Received Signal Strength Indicator}
\newacronym{snr}{SNR}{Signal to Noise Ratio}
\newacronym{sir}{SIR}{Signal to Interference Ratio}
\newacronym{wor}{WOR}{Wake on Radio}
\newacronym{worack}{WOR-ACK}{Wake on Radio Acknowledge}
\newacronym{js}{JS}{Join Server}
\newacronym{as}{AS}{Application Server}
\section{Introduction}

In recent years, the \gls{iot} has revolutionized various sectors, from smart cities to industrial automation, by enabling interconnected devices to communicate seamlessly. At the heart of this transformation lies the need for communication protocols that can achieve wide-area coverage, low power consumption, and scalability to support the ever-increasing number of \gls{iot} devices. LoRa technology, together with the LoRaWAN protocol, has emerged as a leading solution in this context, offering an efficient and cost-effective approach to \glspl{lpwan}~\cite{cuozzo2024support, marini2021lorawansim}.

Traditionally, LoRaWAN has operated primarily in the sub-GHz frequency spectra, such as EU868 MHz, which provide excellent coverage and energy efficiency but are constrained by strict duty cycle regulations and limited data rates. In 2017, the introduction of LoRa chipsets operating at 2.4 GHz aimed to address these limitations. Indeed, the 2.4 GHz spectrum enables higher data rates and no duty cycle restrictions, whereas it also poses challenges related to increased path loss and severe interference. However, the recent incorporation of multi-hop communication has opened new avenues for extending network coverage and improving performance, particularly in challenging environments.

In this paper, we aim to further enhance the performance and capabilities of LoRaWAN by investigating a novel approach that combines the benefits of the sub-GHz and 2.4 GHz spectra, along with the recent introduction of \glspl{rl}, while maintaining adherence to the LoRaWAN standard. Specifically, we propose a multi-band and multi-hop LoRaWAN network retaining \glspl{gw} and \glspl{ed} operating within the sub-GHz frequency, while introducing multi-band \glspl{rl} capable of receiving data from 2.4 GHz \glspl{ed} and forwarding it to sub-GHz \glspl{gw}. This enables the integration of 2.4 GHz into a fully functional LoRaWAN network with minimal modifications to the core architecture. 
Leveraging the findings of our previous papers~\cite{marini2023comparative, cuozzo2024support} and building upon our open-source and standard-compliant simulator \cite{marini2021lorawansim}, we evaluate the network performance of the proposed solution under realistic deployment scenarios. The comprehensive numerical results demonstrate substantial improvements compared to standard single-band and single-hop LoRaWAN networks, including a benchmark LoRaWAN network fully operating at 2.4 GHz, which is motivated by our proposed roadmap \cite{cuozzo2024support}) even if not yet standardized. Through this study, we provide valuable insights into the trade-offs associated with this hybrid architecture, demonstrating its significant potential to revolutionize LPWANs and address the evolving demands of IoT applications.

The structure of the paper is as follows. In Sec.~\ref{sec:technology}, we provide an overview of LoRa and LoRaWAN, including their technological foundations and a literature review. Sec.~\ref{sec:system_model} introduces the system model, detailing the deployment, channel, and traffic assumptions used in our study. Sec.~\ref{sec:approach} describes the proposed multi-band, multi-hop approach, outlining all the different features and mechanisms. In Sec.~\ref{sec:results}, we define the simulation parameters, performance metrics and present the numerical results, highlighting key trends and trade-offs. Finally, Sec.~\ref{sec:conclusions} concludes the paper, summarizing our findings and discussing potential directions for future research.
\section{Technology Background and State of the Art}
\label{sec:technology}

\subsection{LoRa}
\label{sec:lora}

LoRa is a proprietary spread-spectrum modulation technique patented by Semtech~\cite{pasolini2021lora}. It utilizes $M$ distinct chirp signals that span a specific frequency interval, hereafter referred to as \gls{bw} in line with LoRa's terminology. Specifically, $M = 2^{\mathrm{SF}}$, where the \gls{sf} is an integer that determines the temporal evolution of each chirp. Increasing the \gls{sf} results in a longer transmission time (also known as \gls{toa}) but reduces the receiver's sensitivity.
The \gls{toa}, without loss of generality, can be expressed as $\mathrm{ToA} = \frac{2^{\mathrm{SF}}}{\mathrm{BW}}N_{\rm S}$, where \gls{sf} is the spreading factor, \gls{bw} is the bandwidth, and $N_{\rm S}$ represents the number of symbols per symbol time. This value depends on the \gls{sf} and the specific frequency spectrum under consideration. For clarity, throughout this paper, the term \emph{spectrum} refers to the entire frequency range allocated for LoRa transmissions (e.g., the \gls{ism} band), while \emph{channel} denotes a subset of that range. Meanwhile, \gls{bw}, as previously defined, represents the interval over which chirps sweep. 
LoRa also incorporates \gls{fec} mechanisms to balance transmission reliability and time efficiency. Specifically, Hamming codes are employed, with the \gls{cr} determining the number of additional \gls{crc} bits included in a LoRa frame~\cite{loramodulation2015}. A LoRa frame consists of a preamble (for detection and synchronization), an optional header, and the payload containing the actual data.

\subsection{LoRaWAN}

\begin{figure}[!t]
\centering
\includegraphics[width=0.95\textwidth]{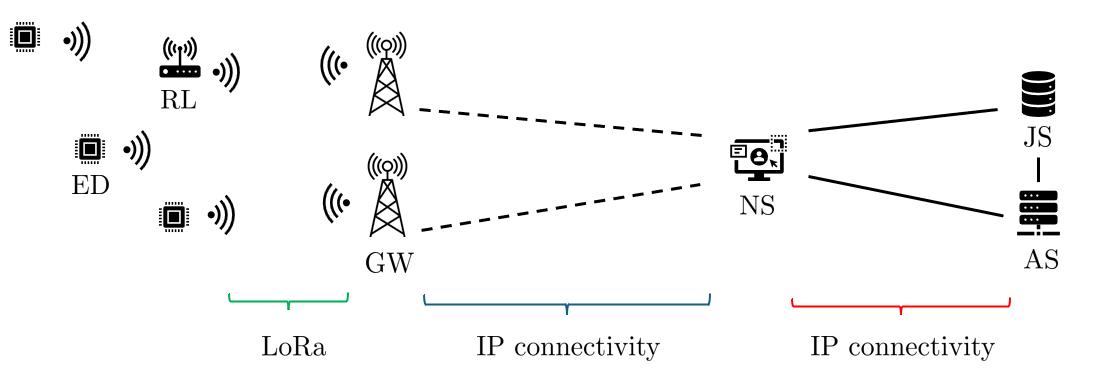}
\caption{Pictorial representation of a LoRaWAN network architecture.}
\label{fig:lorawan_architecture}
\end{figure}

The LoRaWAN protocol, introduced by the LoRa Alliance, builds upon the LoRa physical layer to implement higher-layer functionalities. In particular, Fig.~\ref{fig:lorawan_architecture} represents a block scheme of a LoRaWAN network. \glspl{ed} communicate with \glspl{gw} via LoRa, while the \glspl{gw} act as forwarding entity to the \gls{ns} using backhaul connections like Wi-Fi, 4G/5G, or Ethernet. To maintain clarity and simplicity throughout the text, we will avoid delving into the specific components of the \gls{ns}, such as the home, serving, and forwarding \gls{ns}. However, we remark that the \gls{ns} may also interface with the \gls{js} and \gls{as} to handle critical functions like security, authentication, and other relevant operational aspects. Unlike conventional network architectures, \glspl{ed} are not directly associated with specific \glspl{gw} but rather with the \gls{ns}. Consequently, all \glspl{gw} that receive frames from a given \gls{ed} forward it to the \gls{ns}, which is responsible for generating the corresponding \glspl{ack} in case of confirmed mode.

Recently, LoRa Alliance introduced the concept of \gls{rl}~\cite{lorarelay}, which is a feature designed to extend the network's coverage by enabling certain \glspl{ed} to act as intermediate nodes between other \glspl{ed} and the \gls{gw}. In this mode, a \gls{rl}-enabled \gls{ed} receives uplink messages from nearby \glspl{ed} that are out of direct reach of a \gls{gw} and forwards them towards the \gls{gw} using its own uplink capabilities. This functionality is beneficial in scenarios with sparse gateway deployments or challenging propagation environments, such as remote or indoor locations. \gls{rl} devices must adhere to strict timing and power constraints to ensure compatibility with the LoRaWAN protocol, and their use typically requires careful network planning to avoid interference or excessive energy consumption. This feature enhances the flexibility and scalability of LoRaWAN deployments, especially in hard-to-reach areas.

LoRaWAN relies on the ALOHA protocol for medium access. In uplink communications, \glspl{ed} and \glspl{rl} randomly select an available frequency channel and transmit a LoRa frame whenever a new payload is ready. The transmitted frame is received by all reachable \glspl{gw}, and the \gls{ns} handles duplicate elimination (i.e., removing identical frames forwarded by multiple \glspl{gw}).
On the other hand, LoRaWAN defines three operational classes for downlink communication. In class A, after an uplink transmission, \glspl{ed} open two short receive windows for downlink communication (\gls{rx1} and \gls{rx2}) at fixed intervals of 1 second and 2 seconds, respectively. The \gls{ns} selects the \gls{gw} for the downlink based on the quality of received uplink frames. In class B, devices can open additional receive slots at scheduled times, synchronized through periodic beacons broadcast by \glspl{gw}. Finally, in class C, devices maintain nearly continuous receive windows, except during uplink transmissions, at the expense of increased power consumption. LoRaWAN also supports two different transmission modes, which are called \textit{confirmed} mode (the \gls{ed} requires an \gls{ack} from the \gls{ns} through the \gls{gw} after an uplink transmission) and \textit{unconfirmed} mode (the \gls{ed} transmits data without expecting an \gls{ack} from the \gls{ns}, leaving the \gls{ed} unaware of whether the packet was successfully received).

\subsection{Frequency spectrum}

LoRaWAN can operate across different frequency spectra, each presenting unique trade-offs. Traditionally, the protocol is used in sub-GHz bands (e.g., EU868 and US915 for Europe and the US, respectively~\cite{loraregional}). In these spectra, \glspl{ed} must support default channels (e.g., three in EU868) and comply with duty cycle limitations (e.g., 1\% on default channels) to minimize interference. In 2017, Semtech introduced LoRa chipsets capable of operating in the 2.4 GHz spectrum~\cite{marini2023comparative}. This higher frequency spectrum enables wider channels and higher data rates, but at the cost of increased path loss and lower maximum transmit power. Unlike sub-GHz bands, the 2.4 GHz spectrum does not impose duty cycle limitations or predefined channels, offering flexibility for potential future standardization by the LoRa Alliance.

\subsection{State of the Art}

While LoRaWAN in the sub-GHz band has been extensively studied \cite{marini2021lorawansim}, research on LoRa at 2.4 GHz has primarily focused on interference \cite{polak2020performance} and ranging applications \cite{derevianckine2023opportunities}, with limited attention given to network-level considerations. Among the few works addressing this gap, \cite{cuozzo2021LoRa} analyzes a custom on-demand \gls{mac} protocol for industrial machines leveraging energy harvesting on LoRa at 2.4 GHz, \cite{ferretti2022lora} explores a railway LoRa network that integrates point-to-point communications at 2.4 GHz with LoRaWAN transmissions at EU868 MHz, while \cite{rosler2023opportunistic} demonstrates the benefits of opportunistic routing in a 2.4 GHz LoRa mesh network through system-level simulations. 

To the best of the authors' knowledge, research on LoRaWAN at 2.4 GHz narrows further to \cite{schappacher2021implementation}, which proposes a hybrid approach combining LoRaWAN and IEEE 802.15.4, and \cite{masek2022performance}, which evaluates the success rate of LoRaWAN across EU868, US915, and 2.4 GHz spectra using network simulations. However, these works are highly specific to a few parameters and scenarios.

To address these shortcomings, our earlier work \cite{marini2023comparative} utilized the previously developed network simulator in \cite{marini2021lorawansim} to perform a comprehensive comparison of LoRaWAN in the EU868 and 2.4 GHz spectra. Building on these promising results, in \cite{cuozzo2024support} we proposed a roadmap outlining the integration of the 2.4 GHz band into the current LoRaWAN standard, dividing the process into three main stages based on implementation complexity. In this paper, we take the next step by presenting numerical results evaluating the performance of multi-hopping in a multi-band LoRaWAN network (i.e., working at both EU868 and 2.4 GHz), corresponding to the final stage of our proposed roadmap. We compare the proposed approach with two benchmarks: a LoRaWAN network operating at EU868 MHz, representing the state-of-the-art solution (stage 0 of our roadmap), and LoRaWAN at 2.4 GHz (stage 1, option 1.1 of our roadmap).
%%%%%%%%%%%%%%%%%%%%%%%%%%%%%%%%%%%%%%%%%%%%%%%%%%%%%%%%%%%%%%%%%%
\section{System model}
\label{sec:system_model}
%%%%%%%%%%%%%%%%%%%%%%%%%%%%%%%%%%%%%%%%%%%%%%%%%%%%%%%%%%%%%%%%%%
\subsection{Deployment model}
We consider a scenario where $N$ \glspl{ed} and $R$ \glspl{rl} are randomly and uniformly distributed within a square area of size $A_{\rm L}$. A single \gls{gw} is positioned at the center of the square area. We assume an urban context, meaning that both \glspl{ed} and \glspl{rl} may be located either inside or outside buildings. Buildings are modeled as squares with side length $A_{\rm S}$, and their centers are distributed across the area in a grid pattern with a pitch $A_{\rm P} \geq A_{\rm S}$.

%%%%%%%%%%%%%%%%%%%%%%%%%%%%%%%%%%%%%%%%%%%%%%%%%%%%%%%%%%%%%%%%%%
\subsection{Channel model}
\label{sec:channel_model}
The channel model utilized in this work aligns with the \gls{3gpp} TR 38.901~\cite{3gpp38901} specifications, which apply to frequencies between 0.5 and 100 GHz. Specifically, we adopt the \gls{3gpp} Urban Macro (UMa) propagation model. Consequently, the received power is calculated as follows:

\begin{equation} \label{eq:received_power} 
P_{\rm R} [dBm] = P_{\rm T}[dBm] + G_{\rm T}^{\rm A}[dB] + G_{\rm R}^{\rm A}[dB] - PL[dB], 
\end{equation}

where $P_{\rm T}$ represents the transmission power, $G_{\rm T}^{\rm A}$ and $G_{\rm R}^{\rm A}$ denote the transmit and receive antenna gains, respectively, and $PL$ is the path loss, computed in accordance with \cite{3gpp38901}, which varies based on the \gls{los}/\gls{nlos} condition and geometric parameters (see Tables 7.4.1-1 and 7.4.2-1 in \cite{3gpp38901}). Using eq. \eqref{eq:received_power}, a link (e.g., \gls{gw}-\gls{rl} or \gls{rl}-\gls{ed}) is considered to have coverage if the received power $P_{\rm R}$ exceeds the receiver sensitivity $R_{\rm S}$. The sensitivity $R_{\rm S}$ depends on the \gls{sf} and frequency band being used, with specific values provided in \cite{SX1280, SX1272}.

In addition to coverage requirements, successful decoding of a LoRa frame during a collision necessitates that the \gls{sir} meets or exceeds a defined capture threshold $\gamma$:
\begin{equation} \label{eq:sir} SIR = \frac{P_{\rm R}}{\sum_{i}P_{\rm R_i}} \geq \gamma, \end{equation}
where $P_{\rm R}$ is the desired received power, $P_{\rm R_i}$ denotes the power from the $i$-th interfering LoRa device, and $\gamma$ is specified in Table 6 of \cite{marini2021lorawansim}. Notably, inter-\gls{sf} interference is disregarded, considering only devices transmitting with the same \gls{sf} as potential interferers.

For LoRaWAN operating in the 2.4 GHz band, additional \glspl{sf} (\gls{sf}5 and \gls{sf}6) are available, which are not covered in Table 6 of~\cite{marini2021lorawansim}. To address this, we conservatively assign them the same $\gamma$ values as the diagonal elements of the table, ensuring consistency without compromising generality.

%%%%%%%%%%%%%%%%%%%%%%%%%%%%%%%%%%%%%%%%%%%%%%%%%%%%%%%%%%%%%%%%%%
\subsection{Traffic model}
\label{sec:traffic_model}
In this paper, we focus exclusively on uplink communications. \glspl{ed} periodically generate uplink traffic with a fixed interval of $T_{\rm U}$, where each transmission carries a payload of $B_{\rm U}$ bytes. Instead, \glspl{rl} transmit a frame of size $B$, accumulating data received from \glspl{ed} until the maximum payload size $B_{\mathrm{max}}$, allowed by their respective \gls{sf}, is reached. Only then do they initiate transmission to the \gls{gw}. The values of $B_{\mathrm{max}}$ as a function of the \gls{sf} are detailed in Table~\ref{tab:payload_size}. Notably, we consider only Class A devices operating in unconfirmed mode (see Sec.~\ref{sec:technology}).

\begin{table}[]
    \centering
    \begin{tabular}{|c|c|c|c|c|c|c|}
    \hline
    SF & 7 & 8 & 9 & 10 & 11 & 12\\
    \hline
    $B_{\mathrm{max}}$ & 222 & 222 & 115 & 51 & 51 & 51\\
    \hline
    \end{tabular}
    \caption{The maximum allowed payload size as a function of the \gls{sf} [bytes].}
    \label{tab:payload_size}
\end{table}

%%%%%%%%%%%%%%%%%%%%%%%%%%%%%%%%%%%
\section{Proposed Approach}
\label{sec:approach}

\begin{figure}[!t]
\centering
\includegraphics[width=.8\textwidth]{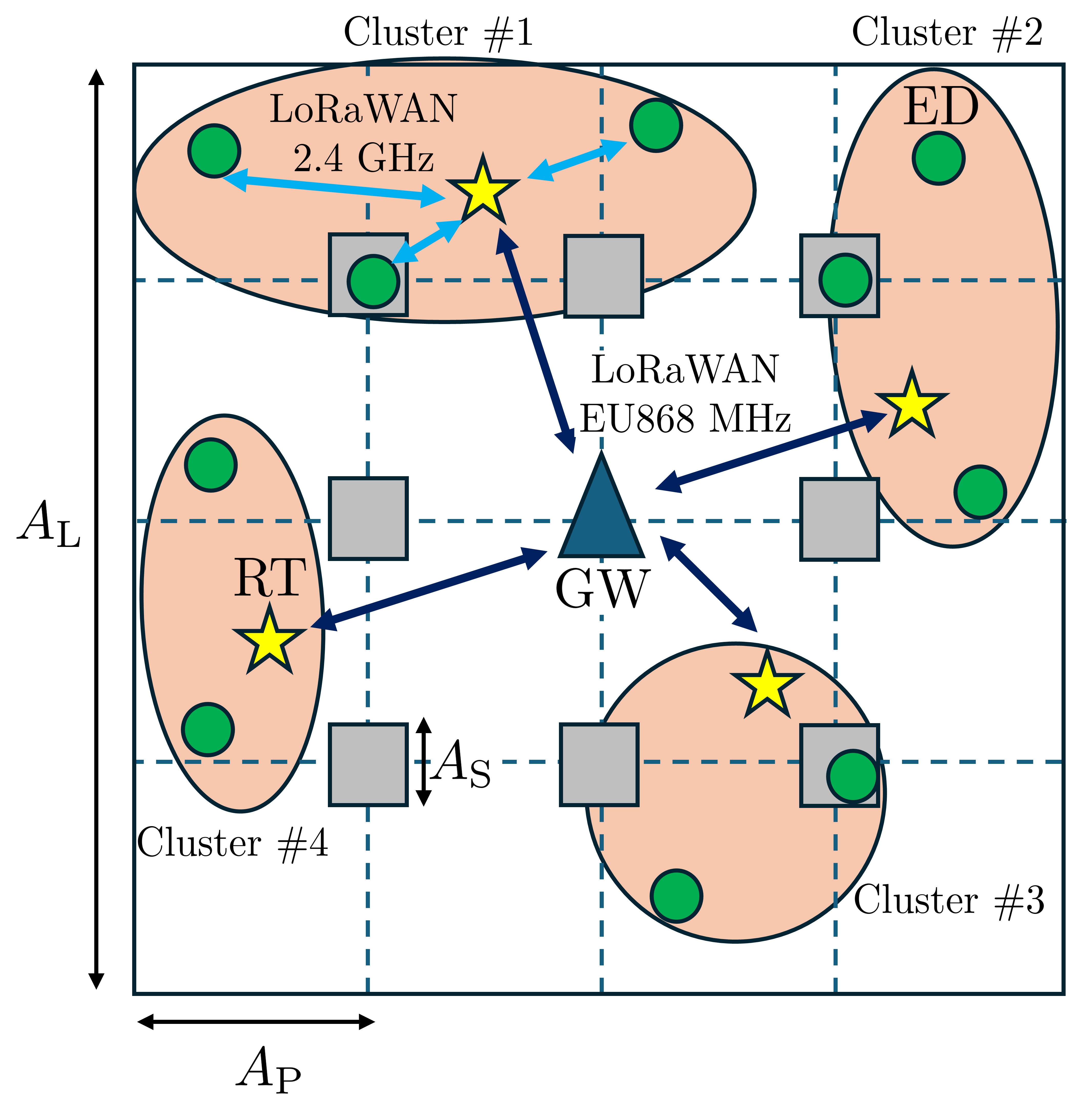}
\caption{Exemplary representation of the proposed network architecture. \glspl{ed} (green circles) communicate with \glspl{rl} (yellow stars) using LoRaWAN at 2.4 GHz, forming clusters where each of them operates on a dedicated radio channel within this spectrum. \glspl{rl} transmit both their own data and that of their associated \glspl{ed} to a central \gls{gw} (blue triangle) via LoRaWAN at EU868 MHz. Links can be in either \gls{los} or \gls{nlos}, depending on obstructions caused by buildings (gray squares).}
\label{fig:network_architecture}
\end{figure}

As already stated, this paper aims to evaluate the network performance of a multi-band and multi-hop LoRaWAN network. In this section, we provide details on multi-band support and on the implementation of multi-hopping.

\subsection{Network architecture}

An example of the proposed network architecture is depicted in Fig.~\ref{fig:network_architecture}\footnote{Notice that hereafter we will neglect the \gls{ns}, \gls{js}, \gls{as} as our focus is on the radio part.}. Each \gls{ed} is equipped with a single 2.4 GHz LoRa transceiver, while \glspl{rl} utilize two LoRa radios, one operating at 2.4 GHz and the other in the EU868 MHz spectrum\footnote{It is worth mentioning that the choice of the EU868 MHz spectrum does not undermine the generality of our approach, as the results are also valid for the other available LoRaWAN frequency portions (e.g., US915 MHz).}. In this setup, \glspl{ed} communicate with \glspl{rl} via LoRaWAN in the 2.4 GHz spectrum \cite{marini2023comparative}, while the \glspl{rl}-to-\gls{gw} links employ LoRaWAN in the EU868 MHz band. \glspl{rl} transmits LoRa frames containing \gls{mac} payloads from both the \glspl{rl} themselves and the \glspl{ed}. 

\subsection{Cluster definition}

Each \gls{ed} communicates with a single \gls{rl}, selected based on the highest received power level $P_{\rm R}$ in the uplink (see eq.~\eqref{eq:received_power}). This can be accomplished in practical implementations by suitably extending the relay management procedures recently introduced by the LoRa Alliance~\cite{lorarelay}\footnote{Specifically, we believe that a four-way handshaking process is required, wherein the \gls{worack} frame includes the received power level of the \gls{wor} frame. This frame is transmitted on a default channel used by all \glspl{rl} for channel sounding. Following this, two additional messages would be exchanged: one from the \gls{ed} to the \gls{rl}, specifying the selected \gls{rl}, and the other from the \gls{rl} to the \gls{ed}, indicating the channel to be utilized.}. The sub-network formed by one \gls{rl} and the associated \glspl{ed} will be referred to as a \emph{cluster} (see Fig. \ref{fig:network_architecture}). Inter-cluster interference is eliminated by assigning a unique channel to each cluster within the 2.4 GHz spectrum. According to recent guidelines from Semtech \cite{semtechphyism}, LoRa 2.4 GHz \glspl{ed} must operate within the 2400 to 2480 MHz frequency band and support a minimum of 16 channels. Consequently, we will adhere to this limit for the number of \glspl{rl}. 

\subsection{Communication model}

The data rate for \glspl{rl} is determined using the well-known \gls{adr} algorithm proposed by Semtech \cite{marini2021adr}. However, in our analysis, the data rate is computed only once and remains fixed throughout the entire simulation run, aligning with the stationarity of the scenario under consideration. Furthermore, we tweaked the algorithm to ensure that \glspl{rl} are assigned with different \glspl{sf} when possible, thereby leveraging their quasi-orthogonality property \cite{pasolini2021lora}. By checking which \gls{sf} have already been assigned to other \glspl{rl}, if a conflict occurs, the \gls{sf} of one or more \glspl{rl} is adjusted to a higher value, ensuring unique assignments while maintaining connectivity with the \gls{gw}. As detailed in Sec.~\ref{sec:traffic_model}, once the data rate is selected, the \glspl{rl} transmit LoRa frames with \gls{mac} payloads adhering to the $B_{\mathrm{max}}$ values specified in Table~\ref{tab:payload_size}. Notably, this approach implies that the uplink transmission periodicity is governed by the 1\% duty cycle restriction of the EU868 MHz spectrum. %When the \gls{rl} is waiting for the next transmission opportunity, it waits for LoRa frames from its \glspl{ed} and initiates the next transmission only after queuing the maximum payload allowed by its assigned data rate. 
On the other hand, within each cluster, 2.4 GHz \glspl{ed} exploit the \gls{adr} algorithm to select the \gls{sf} to be used according to the performance of the \gls{ed}-\gls{rl} link.

%In this paper, we do not simulate the network formation phase, as our focus lies on evaluating the performance of the network once it is fully operational. Nonetheless, we note that the relay management procedures recently introduced by the LoRa Alliance \cite{lorarelay} could be adapted to support our proposed approach.
%%%%%%%%%%%%%%%%%%%%%%%%%%%%%%%%%%%%%%%%%%%%%%%%%%%%%%%%%%%%%%%%%%
% %%%%%%%%%%%%%%%%%%%%%%%%%%%%%%%%%%%%%%%%%%%%%%%%%%%%%%%%%%%%%%%%%%
\section{Performance Evaluation}
\label{sec:results}

\subsection{Simulation Setup}

\textbf{Configuration}\\
Numerical results have been obtained starting from our open-source and standard-compliant simulator published in~\cite{marini2021lorawansim}, properly modified to consider the context of this paper. All results have been obtained by averaging over 1000 simulation runs of $T$ seconds. As for \glspl{ed} configuration, they have fixed positions and they remain in such locations for the entire duration of the simulation $T$. We consider two radio modules, namely the Semtech SX1272~\cite{SX1272} for the EU868 MHz spectrum and the Semtech SX1280~\cite{SX1280} for the 2.4 GHz one, and the corresponding parameters for simulations are taken from their data sheets. The list of simulation parameters is shown in Table~\ref{tab:simulation_parameters}. 

\begin{table}[!t]
\caption{Simulation parameters.}
\centering
\label{tab:simulation_parameters}
\begin{tabular}{|c|c|}
\hline
\textbf{Parameter} & \textbf{Value}\\ \hline
$G_{\rm ED}^{\rm A}$ & 0 dB \\ \hline
$P_{\rm ED}^{\rm T}$ & 12.5 dBm \\ \hline

$P_{\rm RL}^{\rm T}$ & 16 dBm \\ \hline
$G_{\rm RL}^{\rm A}$ & 0 dB \\ \hline

$G_{\rm GW}^{\rm A}$ & 0 dB \\ \hline

BW & 125 (@EU868 MHz) kHz \\ & 203 (@2.4 GHz) kHz \\ \hline
CR & 4/5 \\ \hline

$B_{\rm U}$ & 10 B \\ \hline
$T_{\rm U}$ & 1 s \\ \hline

$T$ & 300 s \\ \hline
\end{tabular}
\end{table}

\textbf{Benchmarking the proposal}\\
%Although the LoRaWAN standard has not yet been updated to fully accommodate this technology, in this paper, we consider \emph{LoRaWAN at 2.4 GHz} as one of the potential LoRaWAN variations. This consideration is made with the anticipation that it could become a future standardization step by the LoRa Alliance, as we have proposed in \cite{cuozzo2024support}. Therefore, the results shown in the following consider two benchmarks related to a LoRaWAN-compliant EU868 MHz network and a LoRaWAN system operating at 2.4 GHz, as well as the paper proposal:

Numerical results consider the following three network architectures:

\begin{itemize}
    \item \textbf{sub-GHz only}, a LoRaWAN network where \glspl{ed} are working in the EU868 MHz band. It constitutes the first benchmark (i.e., stage 0 of the roadmap proposed in \cite{cuozzo2024support});
    \item \textbf{2.4 GHz only}, a LoRaWAN network where \glspl{ed} are working in the 2.4 GHz band and exploit the full LoRaWAN protocol stack to communicate towards a 2.4 GHz \gls{gw}. While this configuration is not currently defined within the official LoRaWAN standard, it serves as a valuable benchmark for comparison (i.e., stage 1, option 1.1, of the roadmap proposed in \cite{cuozzo2024support}, and the object of the analysis of \cite{marini2023comparative});
    \item \textbf{Proposal}, a LoRaWAN network with the introduction of \glspl{rl} to connect 2.4 GHz \glspl{ed} (see Sec. \ref{sec:approach}). 
\end{itemize}

\textbf{Performance metrics}\\
In the following, results are presented in terms of network throughput $S$, defined as the number of LoRa frames correctly received by the \gls{ns} (via \glspl{gw}):

\begin{equation}
    S = \frac{\sum_{i=1}^{R}8 \; M_i \; B_i}{T} \quad \mathrm{[bits/s]}, 
\end{equation}

where $R$ is the number of \glspl{rl} in the network, $M_i$ denotes the number of uplink LoRa frames of size $B_i$ correctly received by the \gls{gw} from \gls{rl} $i$.

In this context, two main factors can lead to a missed reception at the \gls{gw}: noise and interference. Regarding noise, the interplay between the selected \gls{sf}, \gls{bw}, \gls{cr} and $P_{\rm R}$ may prevent successful decoding, even when users are within the coverage of at least one \gls{gw}. As for interference, the \gls{gw}'s ability to decode a signal depends on the power levels of both the desired and interfering signals, as well as the number of interfering transmissions using the same \gls{sf}. These factors determine whether the \gls{sir} at the \gls{gw} exceeds the capture threshold $\gamma$ (see Sec.~\ref{sec:system_model}).

Additionally, we analyze energy consumption suffered by \glspl{ed} during the entire simulation duration. Energy consumption is computed following the model as in~\cite{marini2021lorawansim} and previously used and described in~\cite{marini2023comparative}.

\subsection{Numerical Results}

\begin{figure}[t]
    \centering
    \resizebox{0.99\columnwidth}{!}{
        \begin{tikzpicture}
\begin{axis}[
    width=\textwidth,
    height=\textwidth,
    grid=major,
    grid style=dashed,
    xlabel={\textit{N}, Number of EDs},
    ylabel={\textit{S}, Network Throughput [bit/s]},
    legend pos=north west,
    legend style={fill opacity=0.8, font=\scriptsize},
    nodes={scale=0.8, transform shape},
    tick label style={font=\footnotesize},
]

% Data points for each line
\addplot[mark=square*, mark options={fill=yellow}, yellow, thick, dashed] 
coordinates {(1,160) (50,2357.6) (100,3211.0) (250,4172.9) (500,4624.5)};
\addlegendentry{2.4 GHz only}

\addplot[mark=square*, mark options={fill=black}, black, thick] 
coordinates {(1,35.52) (50,1326.1) (100,1533.3) (250,1669.4) (500,1728.6)};
\addlegendentry{Proposal, R=1}

\addplot[mark=square*, mark options={fill=green}, green, thick] 
coordinates {(1,35.52) (50,1628.6) (100,2391.7) (250,3025.1) (500,3285.6)};
\addlegendentry{Proposal, R=2}

\addplot[mark=square*, mark options={fill=orange}, orange, thick] 
coordinates {(1,35.52) (50,1680.0) (100,2450.0) (250,3600.0) (500,4000.0)};
\addlegendentry{Proposal, R=3}

\addplot[mark=square*, mark options={fill=blue}, blue, thick] 
coordinates {(1,35.52) (50,1745.8) (100,2514.9) (250,4066.3) (500,4422.8)};
\addlegendentry{Proposal, R=4}

\addplot[mark=square*, mark options={fill=red}, red, thick] 
coordinates {(1,35.52) (50,1832.8) (100,2705.6) (250,4450.4) (500,4732.4)};
\addlegendentry{Proposal, R=5}

\addplot[mark=square*, mark options={fill=magenta}, magenta, thick, dashed] 
coordinates {(1,80) (50,224) (100,100) (250,80) (500,76.9)};
\addlegendentry{sub-GHz only}

\end{axis}
\end{tikzpicture}
    }
    \caption{Network throughput as a function of the number of EDs, the number of RLs, and for the two benchmark cases. We set $A_{\rm L}=1$ km.}
    \label{fig:SvEDs_A=1}
\end{figure}
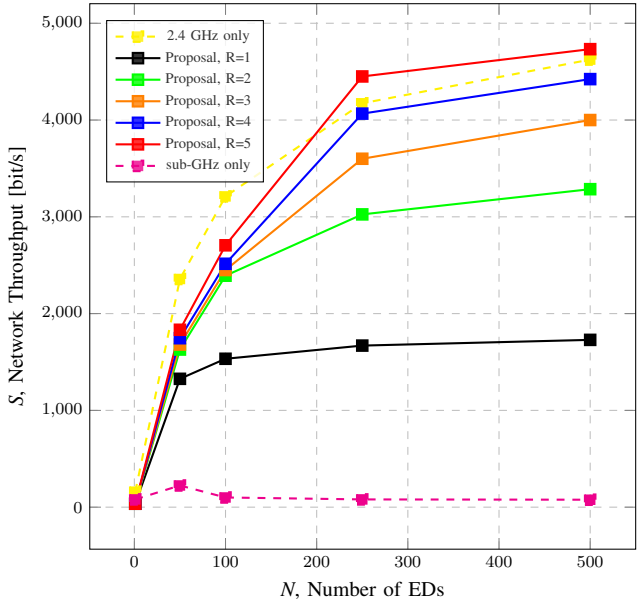

\begin{figure}[t]
    \centering
    \resizebox{0.99\columnwidth}{!}{
        \begin{tikzpicture}
\begin{axis}[
    width=\textwidth,
    height=\textwidth,
    grid=major,
    grid style=dashed,
    xlabel={\textit{N}, Number of EDs},
    ylabel={\textit{S}, Network Throughput [bit/s]},
    legend pos=north west,
    legend style={fill opacity=0.8, font=\scriptsize},
    nodes={scale=0.8, transform shape},
    tick label style={font=\footnotesize},
]

% Data points for each line
\addplot[mark=square*, mark options={fill=yellow}, yellow, thick, dashed] 
coordinates {(1,64) (50,20) (100,39.732) (250,144.8) (500,298.08)};
\addlegendentry{2.4 GHz only}

\addplot[mark=square*, mark options={fill=black}, black, thick] 
coordinates {(1,120) (50,301.92) (100,710.4) (250,1409) (500,1580.6)};
\addlegendentry{Proposal, R=1}

\addplot[mark=square*, mark options={fill=green}, green, thick] 
coordinates {(1,120) (50,497.21) (100,900.3) (250,2200) (500,2800)};
\addlegendentry{Proposal, R=2}

\addplot[mark=square*, mark options={fill=orange}, orange, thick] 
coordinates {(1,120) (50,692.5) (100,1332) (250,2515) (500,3042.7)};
\addlegendentry{Proposal, R=3}

\addplot[mark=square*, mark options={fill=blue}, blue, thick] 
coordinates {(1,120) (50,935.32) (100,1562.7) (250,3625.7) (500,3754)};
\addlegendentry{Proposal, R=4}

\addplot[mark=square*, mark options={fill=red}, red, thick] 
coordinates {(1,120) (50,934.52) (100,1831.9) (250,3776.9) (500,4374.8)};
\addlegendentry{Proposal, R=5}

\addplot[mark=square*, mark options={fill=magenta}, magenta, thick, dashed] 
coordinates {(1,80) (50,838.612) (100,1128) (250,1470.1) (500,1236.5)};
\addlegendentry{sub-GHz only}

\end{axis}
\end{tikzpicture}
    }
    \caption{Network throughput as a function of the number of EDs, the number of RLs, and for the two benchmark cases. We set $A_{\rm L}=5$ km.}
    \label{fig:SvEDs_A=5}
\end{figure}
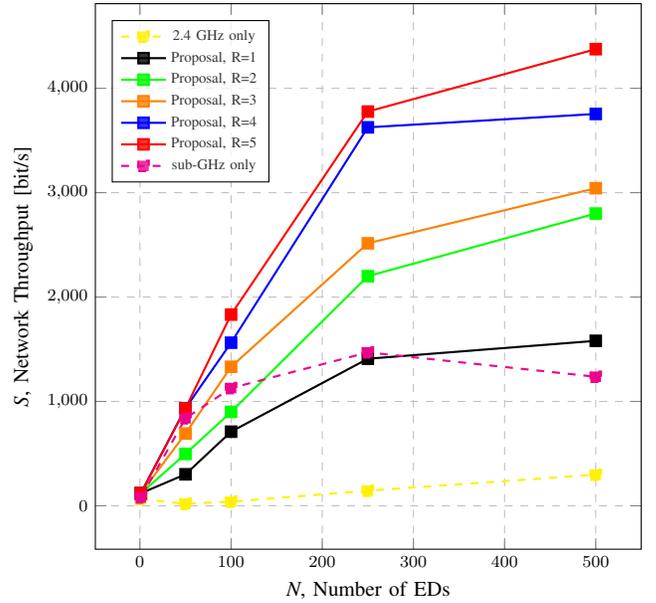

Fig.~\ref{fig:SvEDs_A=1} and ~\ref{fig:SvEDs_A=5} show the network throughput $S$ as a function of the number of \glspl{ed}, $N$, the number of \glspl{rl}, $R$, and for two area dimensions ($A_{\rm L}=1$ and $A_{\rm L}=5$, respectively). We compare our proposal with the two benchmark cases, as described before.
As a general trend, the network throughput increases with the number of \glspl{ed}. Nonetheless, as $N$ increases heavily, the likelihood of collisions rises, potentially limiting further gains in overall network throughput (as can be seen for $N=50$ in the sub-GHz case). This is due to the usual behavior of the ALOHA protocol, upon which LoRaWAN is based. 
As can be seen, after introducing a sufficient number of \glspl{rl}, the proposal is able to overcome both benchmarks (up to 97\%). This improvement is attributed to the creation of a corresponding number of clusters in the network, with each cluster operating on a dedicated channel. This design choice allows to reduce the number of collisions between 2.4 GHz \glspl{ed}, leading to enhanced overall performance. 

In essence, the addition of \glspl{rl} not only alleviates congestion but also optimizes resource allocation, allowing for a more efficient communication environment, especially as the number of \glspl{ed} increases. As expected, the performance gains achieved by our multi-band and multi-hop LoRaWAN network are influenced by the area dimension. In Fig.~\ref{fig:SvEDs_A=1}, where the area is relatively small, the 2.4 GHz LoRaWAN version demonstrates good overall coverage. Consequently, our proposed approach offers a noticeable performance improvement only with a higher number of \glspl{rl} ($R=$ 5). However, the significance of \glspl{rl} becomes more pronounced as the area dimension increases (see Fig.~\ref{fig:SvEDs_A=5}). In this larger scenario, employing just two \glspl{rl} when $N > 100$ is sufficient to achieve an enhancement in network throughput.

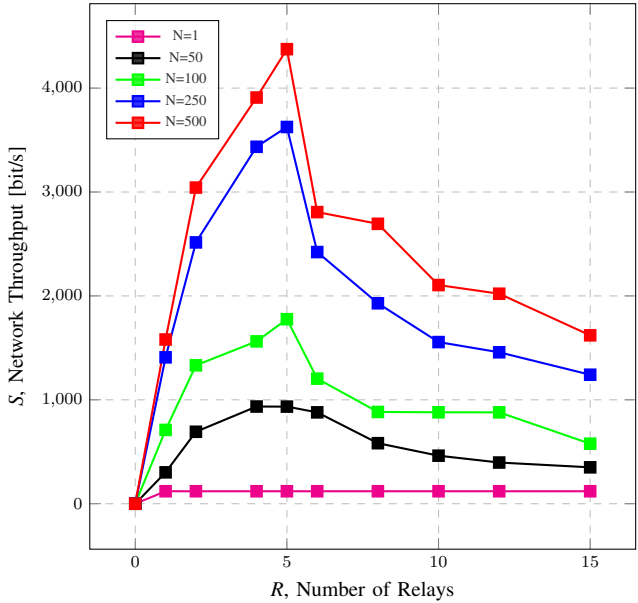
\begin{figure}[t]
    \centering
    \resizebox{0.99\columnwidth}{!}{
        \begin{tikzpicture}
\begin{axis}[
    width=\textwidth,
    height=\textwidth,
    grid=major,
    grid style=dashed,
    xlabel={\textit{R}, Number of Relays},
    ylabel={\textit{S}, Network Throughput [bit/s]},
    legend pos=north west,
    legend style={fill opacity=0.8, font=\scriptsize},
    nodes={scale=0.8, transform shape},
    tick label style={font=\footnotesize},
]

% Data points for each line
\addplot[mark=square*, mark options={fill=magenta}, magenta, thick] 
coordinates {(0,0) (1,120) (2,120) (4,120) (5,120) (6,120) (8,120) (10,120) (12,120) (15,120)};
\addlegendentry{N=1}

\addplot[mark=square*, mark options={fill=black}, black, thick] 
coordinates {(0,0) (1,301.92) (2,692.5) (4,935.32) (5,934.52) (6,879.35) (8,582.59) (10,462.35) (12,396.5167) (15,350.2632)};
\addlegendentry{N=50}

\addplot[mark=square*, mark options={fill=green}, green, thick] 
coordinates {(0,0) (1,710.4) (2,1332) (4,1562.7) (5,1775.8) (6,1203.7) (8,883.155) (10,880.1) (12,879.35) (15,577.64)};
\addlegendentry{N=100}

\addplot[mark=square*, mark options={fill=blue}, blue, thick] 
coordinates {(0,0) (1,1409) (2,2515) (4,3435.3) (5,3625.7) (6,2422) (8,1929) (10,1555.7) (12,1457.5) (15,1241.4)};
\addlegendentry{N=250}

\addplot[mark=square*, mark options={fill=red}, red, thick] 
coordinates {(0,0) (1,1580.6) (2,3042.7) (4,3909.3) (5,4374.8) (6,2806.7) (8,2693.9) (10,2104.4) (12,2021) (15,1620.7)};
\addlegendentry{N=500}

\end{axis}
\end{tikzpicture}
    }
    \caption{Network Throughput as a function of the number of RLs, and the number of EDs, with $A_{\rm L}=5$ km.}
    \label{fig:SvRTs_A=5}
\end{figure}

The impact of the number of \glspl{rl} is also analyzed in Fig.~\ref{fig:SvRTs_A=5}, where $S$ is presented as a function of $R$ and for different values of $N$. The curves show a common trend with an optimum reached for $R=5$. This happens because, for a higher number of \glspl{rl}, the \gls{sf} distribution is no longer orthogonal (i.e., some \glspl{rl} are forced to use the same \gls{sf}), leading to inter-cluster interference.

\begin{table}[]
    \centering
    \begin{tabular}{|c|c|c|c|}
    \hline
    \textbf{Configuration} & $\mathbf{N=50}$ & $\mathbf{N=500}$\\ \hline
    EDs @Sub-GHz & 3382.6 & 3395.01 \\ \hline
    EDs @2.4 GHz & 1420.02 & 1487.9 \\ \hline
    EDs @Proposal & 1117.62 &  1139.17 \\ \hline
    \end{tabular}
    \caption{Average energy consumption of \glspl{ed} as a function of the number of \glspl{ed} and the three network configurations [mJ].}
    \label{tab:energy_consumption}
\end{table}

In Table~\ref{tab:energy_consumption}, we analyze the average energy consumption suffered by \glspl{ed}, where we set $R=5$ (i.e., the optimum number of \glspl{rl}) and $A_{\rm L}=5$ km (i.e., the largest area dimension). As can be seen, our proposal outperforms the two benchmarks (up to 67\%) for both the extreme values of $N$. In fact, \glspl{ed} consume less energy due to the possibility of exploiting lower \glspl{sf} for their communication, hence reducing the overall time spent in transmission (i.e., the \gls{toa}).

In general, the introduction of \glspl{rl} shows benefits in terms of network performance, even though it should be highlighted that it requires an implementation cost (i.e., the introduction of \glspl{rl} which should provide radio modules for both spectra).
\section{Conclusions}
\label{sec:conclusions}
This paper proposes a novel hybrid architecture for LoRaWAN networks that leverages the complementary features of the EU868 MHz and 2.4 GHz bands, along with the recent introduction of \gls{rl} functionality in the standard, to enhance network performance while preserving standard compliance. In particular, the network operates in the EU868 MHz band, but it introduces \glspl{rl} that are capable of receiving LoRa frames from 2.4 GHz \glspl{ed} and forwarding them to sub-GHz \glspl{gw}. By leveraging our open-source and standard-compliant network simulator, the numerical results demonstrate an average improvement in network throughput of up to 97\% and energy efficiency of up to 67\% compared to single-band (i.e., EU868 MHz or 2.4 GHz) and single-hop deployments. Additionally, our solution exhibits better scalability with increasing numbers of \glspl{ed} and larger deployment areas. The findings highlight the significant potential of this hybrid approach to revolutionize LPWANs and address the evolving demands of next-generation IoT applications.

\IEEEtriggeratref{0}
\bibliographystyle{IEEEtran}
\bibliography{bibl.bib}
\end{document}